\documentclass[pra,twocolumn,aps,amssymb,footinbib]{revtex4}
\usepackage{amssymb}

\usepackage{amssymb}

\usepackage{graphicx}
\usepackage{amsmath}
\usepackage{times}
\usepackage{color}

\begin{document}

\title{ The role of interactions, tunneling and  harmonic confinement on the adiabatic loading of bosons in an
optical lattice.  }

\author{Ana Maria Rey$^{1,2}$ \footnote{Electronic address: arey@cfa.harvard.edu},  Guido Pupillo$^{1,3}$ and J. V. Porto$^{1}$
}

\affiliation{$^{1}$National Institute of Standards and Technology,
Gaithersburg, MD 20899}

 \affiliation{$^{2}$Institute for Theoretical
Atomic, Molecular and Optical Physics, Harvard-Smithsonian Center of
Astrophysics, Cambridge, MA, 02138.}

 \affiliation{$^{3}$Institute
for Quantum Optics and Quantum Information of the Austrian Academy
of Sciences, 6020 Innsbruck, Austria}

\date{July 11, 2005}

\begin{abstract}
We calculate  entropy-temperature curves for interacting
bosons in unit filled optical lattices for both homogeneous and harmonically trapped situations,  and use them to
understand
how adiabatic changes in the lattice depth affect the temperature of
the system.  In a translationally invariant
lattice, the zero tunneling limit facilitates a rather detailed
analytic description. Unlike the non-interacting bosonic system which is always cooled upon adiabatic
loading for low enough initial temperature, the change in the excitation spectrum induced by interactions
 can lead to heating. Finite tunneling helps to  reduce this heating. Finally,  we study the spatially
 inhomogeneous system confined in a parabolic potential and show that the presence of the trap can significantly reduce the final available temperature, due to the
non-vanishing superfluid component at the edge of the cloud which is
present in  trapped systems.

\end{abstract}
\maketitle
\section{Introduction}

Cold atoms in optical lattices provide a system for realizing interacting many-body systems in essentially
defect free lattices~\cite{Jaksch}, and have been an active area of research in recent years. The strong
interest in this system is due in part to the ability to dynamically control lattice parameters at a
level unavailable in more traditional condensed matter systems. Lattice-based systems are typically
governed by three sets of energy scales: interaction energies $U$, tunneling rates $J$ and the temperature $T$.
In atomic systems, the energies $U$ and $J$ can be controlled by adjusting the lattice, and their values can be
 measured and/or calculated easily.  Unlike condensed matter systems, however, it is experimentally
 difficult to measure very low temperatures, ($k T<~J$, $k T  \leq U$, here $k$ is the Boltzmann constant),
  and the temperature has so far only been
  inferred in a few cases\cite{Paredes, Reischl,GCP, Stoef05, Troyer05}. Absent good thermometers, and
  given the ability to dynamically change the density of states, it is important to understand
   the thermodynamics of  experimentally realistic systems in order to estimate the temperature.

It has been pointed out that loading sufficiently cold, non-interacting atoms into an optical lattice can
 lead to adiabatic cooling,~\cite{Blair,Demler}, but the cooling available in
 a real system will clearly depend on and be limited by interactions. It can also depend on the (typically harmonic)
 trapping potential, which provides an additional energy in the problem, as well as on the finite size of the sample.
  Here, we calculate the entropy of bosons in unit filled optical lattices for homogeneous and trapped  cases.
  We provide good approximate, analytical expressions for the entropy for various cases, including finite
  number effects which allow for comparison of temperatures for adiabatic changes in the lattice.

For translationally invariant lattices at commensurate filling, the reduced density of states
 associated with the gap that appears in the insulating state presents a significant limitation
  to the final temperature when raising the lattice \cite{Reischl}. The presence of the trap,
  and the associated superfluid-like component at the edges can significantly increase the density of states,
   however, allowing for lower final temperatures.

   In this paper we make the assumption of adiabatic loading  and
  thus calculate  the lowest possible final temperature achievable from
   a given initial temperature during the loading process. We realize that to be fully adiabatic might be
    experimentally
challenging, however   our calculations could be used to benchmark
the effect of the loading on the temperature of the atomic sample.

The paper is organized as follows: We start by introducing  the
model Hamiltonian and our notation. In Sec. \ref{hom} we focus on
the translationally invariant case. We first develop analytic
expression for the thermodynamic quantities  in the $J=0$ limit and
 then we use them  to calculate the final temperature of the atomic
sample assuming we start with a dilute weakly interacting BEC,
described using the Bogoliubov approximation. Next we study
how finite size effects and  finite $J$ corrections modify the final
temperature of the sample. In Sec. \ref{para}  we discuss  the
effects of a spatial inhomogeneity induced by an additional
parabolic potential and finally in Sec. \ref{concl} we conclude.

 \section{Bose-Hubbard Hamiltonian }
The Bose-Hubbard ({\it BH}) Hamiltonian describes interacting bosons
in a periodic lattice potential  when the lattice is loaded such
that only the lowest vibrational level of each lattice site is
occupied and tunneling occurs only between nearest-neighbors
\cite{Jaksch}
\begin{equation}
H= - \sum_{\langle \textbf{i},\textbf{j}\rangle
}J_{\textbf{i},\textbf{j}}\hat{a}_\textbf{i}^{\dagger}\hat{a}_{\textbf{j}}
+\frac{U}{2}\sum_{\textbf{j}}\hat{n}_\textbf{j}(\hat{n}_\textbf{j}-1) +V_\textbf{j} \hat{n}_\textbf{j}.\\
\label{EQNBHH}
\end{equation}
Here $\hat{a}_\textbf{j}$ is the bosonic annihilation operator of a
particle at site $\textbf{j}=\{j_x,j_y,j_z\}$,
$\hat{n}_\textbf{j}=\hat{a}_\textbf{j}^{\dagger}\hat{a}_{\textbf{j}}$,
and the sum $\langle \textbf{i},\textbf{j}\rangle$ is over nearest
neighbor sites.  $U$ is the interaction energy   cost for
having two atoms at the same lattice site which is proportional to
the scattering length $a_s$, $V_\textbf{j}$ accounts for any other
external potential such as the parabolic magnetic confinement
present in most of the experiments and $J_{\textbf{i},\textbf{j}}$
is the hopping matrix element between nearest neighboring lattice
sites.

For  sinusoidal  separable lattice potentials  with  depths
$\{V_x,V_y,V_z\}$ in the different directions,  the nearest
neighbors hopping  matrix elements, $\{J_x,J_y,J_z\}$, decrease
exponentially with the lattice depth in the respective direction and
$U$ increases as a power law: $U\propto a_s( V_xV_y V_z
)^{1/4}$\cite{Jaksch}.


\section{Homogeneous lattice  }
\label{hom}

\subsection{Thermodynamic properties in the $J=0$ limit}

In this section we calculate expressions for the thermodynamic properties of  $N$ strongly correlated bosons in
 a spatially  homogeneous
lattice ($V_\textbf{i}=0$), with $M$   sites.
For the case where  $J_{x,y,z}=0$, (relevant for very deep lattices) the entropy can be calculated from
 a straightforward accounting of occupation of Fock states, and is independent of the number of spatial dimensions.
We derive expressions for the entropy
per particle as a function of $M,N, U$ and the temperature $T$, in the thermodynamic limit where $N\to  \infty$
 and $M \to  \infty$, while the filling factor $N/M$ remains constant.

In the  $J_{x,y,z}=0$ limit,  Fock number states are eigenstates of
the Hamiltonian and the partition function $\mathcal{Z}$  can be
written as:

\begin{equation}
\mathcal{Z}(N,M) = \sum_{\{n_r\} }\Omega( n_r) e^{-\beta \sum_r
E_rn_r}, \label{Zpa}
\end{equation}

\noindent where  $\beta=(k T)^{-1}$, $k$ is  the Boltzmann constant.
Here we use the following notation:
\begin{itemize}
  \item The quantum numbers  $n_r$ give the
number of wells with $r$ atoms, $r=0,1,\dots,N$,  in a particular
Fock state of the system. For example for a unit filled lattice the
state $|1,1,\dots,1,1\rangle$ has quantum numbers $n_1=N$ and
$n_{r\neq1}=0$.
  \item $E_r\equiv \frac{U}{2}{r(r-1)}$
 \item The sum is over all different configurations ${\{n_r\} }$
 which satisfy the constrains of having $N$ atoms in $M$ wells:
 \begin{eqnarray}
 \sum_{r=0} n_r&=&M,\\
  \sum_{r=0} r n_r&=&N, \label{conn}
\end{eqnarray}
  \item $\Omega( n_r)$ accounts for the number of  Fock states which
  have the same quantum numbers $n_r$ and thus are degenerate  due to the translational invariance
  of the system:

\begin{equation}
\Omega( n_r)=\left(\begin{array}{c}  M
\\n_0,n_1,\dots,n_N \ \end{array}\right)=\frac{M!}{n_0!n_1!\dots},
\end{equation}

\end{itemize}

\noindent Notice that without the particle number constraint,
Eq.(\ref{conn}) and Eq. (\ref{Zpa}) could be easily evaluated. It would
just be  given by

\begin{eqnarray}
\sum_N \mathcal{Z}(M,N) &=& \sum_{n_0,n_1,\dots}
\frac{M!}{n_0!n_1!\dots} (e^{-\beta E_0})^{n_0}(e^{-\beta
E_1})^{n_1}\dots \notag \\&=&\left (\sum_r e^{-\beta
E_r}\right)^M.\label{Zpan}
\end{eqnarray}

\noindent However, the  constraint of having exactly  $N$ atoms,
Eq.(\ref{conn}), introduces some complication. To evaluate the
constrained sum we follow the  standard procedure and  go from a
Canonical to a grandcanonical formulation of the problem.

Defining the grandcanonical partition function:
\begin{equation}
\it{\Xi}(M)\equiv \sum_{N'} \mathcal{Z}(N',M)e^{\beta \mu N'} =
\left(\sum_r e^{-\beta( E_r- \mu r)}\right)^M,\label{GC}
\end{equation} and using the fact that  $\it{\Xi}(M)$ is a very
sharply peaked function, the sum in Eq.(\ref{GC}) can be evaluated
as the maximum value of the summand multiplied by a width $\Delta
N^*$:
\begin{equation}
\it{\Xi}(M)\approx \mathcal{Z}(N,M)e^{\beta \mu N}\Delta N^*,
\label{eq}
\end{equation}Taking the logarithm of the above equation and neglecting the term
$\ln(\Delta N^*)$, which in the thermodynamic limit is very small
compared to the others ($\Delta N^*\ll~N$), one gets an excellent
approximation for the desired partition function,
$\mathcal{Z}(N'=N,M)$:

\begin{eqnarray}
\ln[\mathcal{Z}(N,M)]&=& -\beta \mu N + \ln[\it{\Xi}(M)].
\label{pgc}
\end{eqnarray}

The parameter $\mu$  has to be chosen to maximize
$\mathcal{Z}(N',M)e^{\beta \mu N'}$ at  $N$. This  leads to the
constraint:
\begin{eqnarray}
g &=&\sum_r r \overline{n}_r,\\\overline{n}_r &=&  \frac{ e^{-\beta
(E_r-\mu r)}}{\sum_s e^{-\beta( E_s-\mu s)}},\label{cons}
\end{eqnarray}
where $g=N/M$ is the filling factor of the lattice,
$\overline{n}_r$ is the mean density of lattice sites with $r$
atoms, and $\mu$ is the chemical potential of the gas.

 From  Eq.(\ref{cons}) and  Eq.(\ref{GC}) one
can  calculate all the thermodynamic properties of the system. In
particular, the entropy  per particle of the system can be expressed
as:

\begin{equation}
S(M,N)=k(-\beta \mu  + \frac{1}{N} \ln[\it{\Xi}(M)] + \beta E),
\label{SS}
\end{equation}

\noindent where $E=1/N\sum_r E_r \overline{n}_r$ is the mean energy
per particle.

\subsubsection{Unit filled lattice $ M=N$}

For  the case $M=N$  it is possible to show that, to an excellent
approximation, the solution of Eq.(\ref{cons})is  given by:

\begin{equation}
 \mu=\frac{U}{2}-\ln[2]\frac{e^{-C \beta U}}{\beta}, \label{muana}
\end{equation}
with $C= 1.432$.  Using this value of $\mu$ in the grandcanonical
partition function one can evaluate all the thermodynamic
quantities.


\begin{itemize}
  \item {\em Low temperature limit $(k T <U)$}
\end{itemize}In the low temperature regime $\mu \simeq U/2$. By
replacing  $\mu=U/2 $ in Eq.(\ref{GC}) one can write  an analytic
expression for $\Xi$ and $E$ (and thus for $S$ ) in terms of
Elliptic Theta functions \cite{AS64}   $\vartheta _{3}\left(
z,q\right)~=1+2\sum_{n=1}^{\infty}q^{n^{2}}\cos \left[ 2nz\right]$:

\begin{eqnarray}
\it{\Xi}(N) = \left[1+ \frac{e^{\beta
U/2}}{2}\left(1+\vartheta_3(0,e^{-\beta U/2})\right)\right]^N,
\label{ana1} \end{eqnarray}
\begin{eqnarray}
&& E =  \frac{ U}{2} \left[\frac{ 2+\vartheta'_3(0,e^{-\beta U/2})}{
2+e^{\beta U/2}[1+\vartheta_3(0,e^{-\beta U/2})]}\right],
\label{ana2}
\end{eqnarray} with $\vartheta'_3(z,q)\equiv \partial \vartheta_3(z,q)/\partial q$. In this low temperature regime one
 can also write an analytic
expression for $\overline{n}_r$
\begin{eqnarray}
\overline{n}_r&=&   \left\{\frac{2 e^{-\beta U/2(r-1)^2}}{
2+e^{\beta U/2}[1+\vartheta_3(0,e^{-\beta U/2})]}\right\}
\end{eqnarray}

\begin{itemize}
  \item {\em High temperature limit $(k T >U)$}
\end{itemize}

\begin{figure}[htbh]
\begin{center}
\leavevmode {\includegraphics[width=3.5 in,height=5.5in]{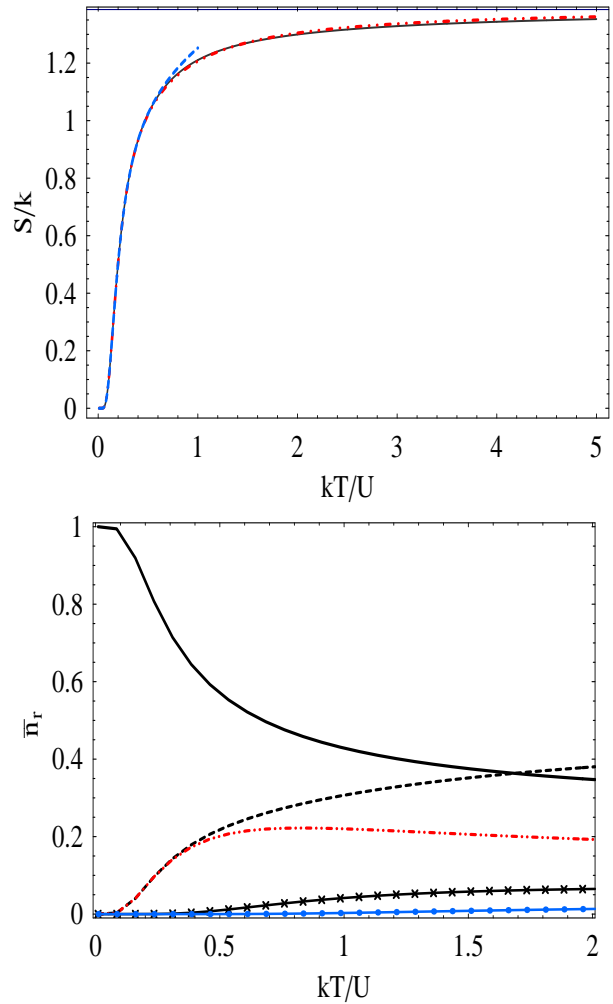}}
\end{center}
\caption{(color online) Top: Entropy per particle  as a function of
the temperature $T$ (in units of U) for a unit filled lattice in the
$J=0$ limit. Dash-dotted (red) line: Eq.(\ref{SS}) calculated using
the numerical solution of Eqs. (\ref{GC}) and (\ref{cons}); Solid
(black) line: entropy calculated using Eq. (\ref{muana}) for the
chemical potential; Dashed (blue) line: Eq.(\ref{SS}) calculated
using the low-temperature analytic solutions: Eqs.(\ref{ana1}) and
(\ref{ana2}). Bottom:  Average occupation number $\overline{n}_r$ as
a function of $T$ (in units of U) . The conventions used are:
 $\overline{n}_1$
(continuous   line), $\overline{n}_0$ ( dashed), $\overline{n}_2$
(dotted-dashed) , $\overline{n}_3$ ( crosses) and $\overline{n}_4$
(dots).}\label{fig1}
\end{figure}

In  the high temperature regime $ {\beta} \mu \simeq-\ln[(1+g)/g]$
which is just $ {\beta} \mu =-\ln2$ for the unit filled case. This
can be easily checked by setting $\beta = 0$ in Eq.(\ref{cons}) and
solving for $\mu$.

For  large temperatures, $\beta \to 0$ , the grandcanonical
partition function and the energy  approach an asymptotic value:
$\ln[\it{\Xi}(M)] \rightarrow M[\ln(1+g) ]$, $ E\rightarrow U g$.
Therefore  the entropy per particle reaches an  asymptotic plateau
$S/k \to \frac{1}{N}\ln\left[\frac{(1+g)^{N+M}}{g^N} \right]\simeq
\frac{\ln[\Omega_o]}{N} $. This plateau can be understood  because $
\Omega_o=\frac{(N+M-1)!}{(M-1)!N!}$ is the number of  all the
possible accessible states to the system in the one-band
approximation (total number of distinct ways to place $N$ bosons in
$M$ wells). It is important to emphasize however, that the one-band
approximation is only valid for $k T\ll E_{gap}$, where $E_{gap}$ is
the energy gap to the second band. { For example, for the case of
${}^{87}$Rb atoms trapped in a cubic lattice potential,
$V_x=V_y=V_z$, $ E_{gap} \geq 10 U $ for lattice depths  $V_x \geq 2
E_R$. Here, $E_R$ is the recoil energy, and $E_R~=h^2/(8 m d^2)$
where $d$ is the lattice constant and $m$ the atoms' mass.} At
higher temperatures the second band starts to become populated and
thus the model breaks down.

In Fig.\ref{fig1} we plot the entropy per particle  as a function of
temperature for a unit filled lattice. The (red) dash-dotted line
corresponds to the  numerical solution of  Eq. (\ref{GC}) and Eq.
(\ref{cons}). The solid  line (barely distinguishable from the
numerical solution) corresponds to entropy calculated using  the
analytic expression of $\mu$ given in Eq. (\ref{muana}). The (blue)
dashed line corresponds to the analytic expression of the entropy
derived for the low temperature regime in terms of Elliptic Theta
functions: Eqs.(\ref{ana1}) and (\ref{ana2}). From the plots one can
see that Eq. (\ref{muana}) is a very good approximation for the
chemical potential.
Also the analytic expression derived for the low temperature regime
reproduces well the numerical solution
 for temperatures $k T<U$.

It is also interesting to note the plateau in the entropy observed
at extremely low temperatures, $ k T<0.05 U$. This plateau is
induced by the gapped excitation spectrum characteristic of an
insulator which exponentially suppresses the population of excited
states at very low temperatures. As we will discuss below the range
of temperature over which the plateau exists is reduced if $J$ is
taken into account.

In the Fig.1 we also show $\overline{n}_r$, the average densities of
sites with $r$  atoms vs temperature calculated using
Eqs.(\ref{muana}) and (\ref{cons}). In particular $\overline{n}_1$
is important because lattice based quantum information proposals
\cite{Jaksch99,Calarco,GAG} rely on having exactly one atom per site
to inizialize the quantum register and population of states with $r
\ne 1$ degrades the fidelity. Specifically we plot $\overline{n}_1$
(solid line), $\overline{n}_0$ (dashed line), $\overline{n}_2$
(dotted-dashed), $\overline{n}_3$ (crosses) and $\overline{n}_4$
(points).

{ In the entropy-plateau region of Fig.1, corresponding to $kT<0.05
U$, particle-hole excitations are exponentially inhibited and thus
$\overline{n}_1$ is almost one.} For temperatures $k T<U /2$,
$\overline{n}_0$ is almost  equal to $\overline{n}_2$, meaning that
only particle-hole excitations are  important. As the temperature
increases, $k T> U /2$, states with three atoms per well start to
become populated and therefore $\overline{n}_0$ becomes greater than
$\overline{n}_2$. The population of states with $r\ge 3$  explains
the break down  of the analytic solution written in terms of
elliptic functions for $k T> U /2$ as  this solution assumes
$\overline{n}_0=\overline{n}_2$. For $k T> 2U $, even states with
$4$ atoms per well become populated and  the fidelity of having unit
filled wells   degrades  to less than $60\%$.

\subsection{Adiabatic Loading }
\label{adilod}

\begin{figure}[tbh]
\begin{center}
\leavevmode {\includegraphics[width=3.5 in,height=5.2
in]{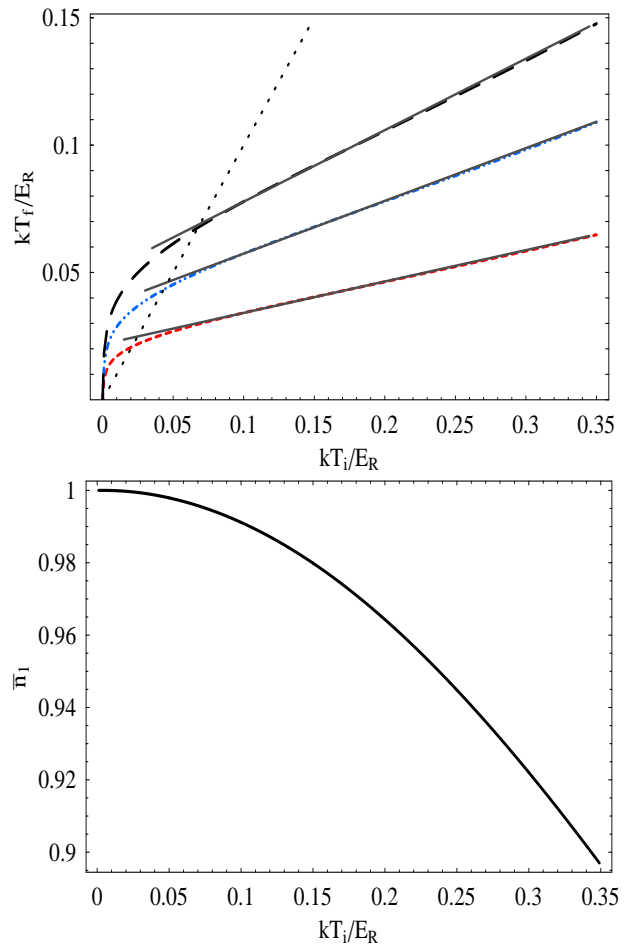}}
\end{center}
\caption{(color online) Top: $T_f$  vs $T_i$ (in units of $E_R$) for
different final lattice depths $V_f$. Here, we assume adiabatic
loading in the limit $J=0$. The dashed(red), dot-dashed(blue), and
long-dashed(black) lines are for $V_f=10, 20$ and $30 E_R$
respectively. The continuous(grey) lines are calculated for the
various lattice depths from Eq.(19). The dotted line is the
identity, $T_f=T_i$. Bottom: Average density of unit filled cells
$\overline{n}_1$ as a function of $T_i$ (in units of $E_R$).
 }\label{fig2}
\end{figure}
 In this section we use the entropy curves derived in
the previous section for the unit filled lattice to calculate how
the temperature of a dilute 3D Bose-Einstein condensate (BEC)
changes as it is adiabatically loaded into a deep optical lattice.
Ideally the adiabatic loading process will transfer a  $T=0$ BEC
into a perfect Mott Insulator (MI), however condensates can not be
created at $T_i=0$ and it is important to know the relation between
final and initial temperatures. Calculations for an ideal bosonic
gas \cite{Blair} demonstrate that for typical temperatures at which
a BEC is created in the laboratory, adiabatically ramping up the
lattice has the desirable effect of cooling the system. On the
other hand, drastic changes in the energy spectrum (the opening up
of a gap) induced by interactions modify this ideal situation
\cite{Reischl} and  in the interacting case atoms can be instead
heated during the loading.

In order to calculate the change in the temperature due to the
loading, we first calculate the entropy as a function of temperature
of a dilute uniform BEC  of $^{87}$Rb atoms  by using Bogoliubov
theory. The Bogoliubov approximation is good for a dilute gas as it
assumes that  quantum fluctuations introduced by interactions are
small and treats them as a perturbation. The quartic term in the
interacting many-body hamiltonian is approximated by a quadratic one
which can be exactly diagonalized \cite{Moelmer,Burnett}. This
procedure yields a quasi-particle excitation spectrum given by
$\epsilon_\textbf{p}=\sqrt{(\epsilon_\textbf{p}^0)^2+2 {\rm u}
n\epsilon_\textbf{p}^0}$. Here
$\epsilon_\textbf{p}^0=\textbf{p}^2/2m$ are single particle
energies, ${\rm u}=4\pi \hbar^2 a_s/m$,  $m$  is the atomic mass and
$n$ is the gas density .

Using this quasi-particle spectrum in the Bose distribution function
of the excited states, $f(\epsilon_\textbf{p}) = [e^{\beta
\epsilon_\textbf{p}} -1]^{-1}$,
 one can evaluate the entropy of the gas given by
\begin{equation}
S|_{V_{x,y,z}=0} = k\sum_\textbf{p} \{\beta \epsilon_\textbf{p}
f(\epsilon_\textbf{p})-\ln[1 -e^{\beta
\epsilon_\textbf{p}}]\}.\label{bog}
\end{equation}Using Eq.(\ref{bog}) we numerically  calculate the entropy of the
system for a given initial temperature $T_i$. Assuming the entropy
during the adiabatic process is kept constant, to evaluate  $T_f$
for a given $T_i$ we solve the equation,
\begin{equation}
S(T_i)|_{V_{x,y,z}=0}=S(T_f)|_{V_{x,y,z}=V_f}.
\end{equation}
We evaluate the right hand side of this equality assuming that the
final lattice depth, $V_f$, is large enough that we can neglect
terms proportional to $J$ in the Hamiltonian. We  use the expression
for the entropy derived in the previous section, Eq.~(12), together
with Eqs.~(14) and (15).

{ The results of these calculations are shown in Fig.\ref{fig2}
where we plot $T_f$ vs $T_i$  for three different final lattice
depths, $V_f/E_R=10$ (dashed  line), 20 (dot-dashed line)
 and 30 (long-dashed  line). 
In the plot both $T_f$ and $T_i$ are given in recoil units $E_R$. As
a reference, the critical BEC temperature for an ideal bosonic gas
(which for a a dilute gas is only slightly affected by interactions)
in recoil units is $kT_c^0\approx 0.67 E_R$.

For $kT_i>0.05 E_R$ the final temperature scales linearly with
$T_i$:
\begin{equation}
kT_f= \frac{U}{3 E_R} \left(kT_i+0.177E_R\right), \label{fit}
\end{equation}In Fig.~\ref{fig2}, Eq.(\ref{fit})  is plotted with a gray  line  for
the various final lattice depths.

In contrast to the non-interacting case, where for $k T_i <0.5 E_R$
the system is always cooled when loading into the lattice
\cite{Blair}, here interactions can heat the atomic sample for low
enough initial temperatures. For reference in Fig.2, we show the
line $T_f=T_i$. One finds a temperature $T^{heat}(V_f)$, (determined
from the intersection of the $T_f=T_i$ line with the other curves)
below which the system heats upon loading into a lattice of depth
$V_f$. From the  linear approximation one finds that $T^{heat}$
increases with $U$ as $kT^{heat}(V_f)\approx 0.177 U
(3-U/E_R)^{-1}$. Because $U$ scales as a power law with the lattice
depth \cite{Jaksch}, a larger $V_f$ implies a larger $T^{heat}(V_f)$
and so a larger heating region.  Note that for the shallowest
lattice in consideration, $V_f/E_R=10$, $kT^{heat}<0.05$ and
therefore
 the linear approximation does not estimate it   accurately.
Fig. 2 also shows a very rapid increase in the temperature close to
$T_i=0$. This drastic increase is due to the low temperature plateau
induced by the gap that opens in the insulating phase.

To quantify the particle-hole excitations and  give an idea of how
far from the target ground state the system is after the loading
process, we also plot $\overline{n}_1$
 vs $T_i$ in the bottom panel of Fig.2 .
In the plot, $\overline{n}_1$ is calculated from Eq.(16).
We found that to a very good approximation
\begin{equation}
\overline{n}_1(T_i)=\left[1-\exp\left(\frac{-3}{2
kT_i/E_R+0.354}\right)\right]^{-1}.
\end{equation}Note that in the $J=0$ limit,
$\overline{n}_1$ depends exclusively on $\beta U$ and thus
  as long as the final lattice depth is large enough to make the $J=0$
approximation valid, $\overline{n}_1$  is independent of the final
lattice depth.
 The exponential suppression of multiple occupied states in the
entropy plateau  explains why even though the final temperature
increases rapidly near $T_i=0$, this is not reflected  as a rapid
decrease  of $\overline{n}_1$.  For the  largest initial temperature
displayed in the plot, $k T_i/E_R\approx T_c^0/2$, the final
temperature reached in units of U is  $kT_f/U\approx 0.17$ and
$\overline{n}_1 \approx 0.9$. Thus, the fidelity of the target state
has been degraded to less than $90\%$. In Fig.1 one also observes
that $\overline{n}_1\approx 0.9$  at $k T_f/U\approx 0.17$   and
that most of the loss of fidelity is due to particle-hole
excitations as $\overline{n}_{r>3}\approx 0$.}

\subsection{Finite size effects }

In recent experiments by loading a BEC into a tight two-dimensional
optical lattice, an array of quasi-one dimensional tubes has been
created \cite{Tolra,Weiss,Paredes,Moritz,Fertig}. The number of
atoms in each tube is of the order of less than $10^2$ and therefore
the assumption of being in the thermodynamic limit is no longer
valid for these systems.

 The thermodynamic limit assumption  used in the previous section allowed
us to derived thermodynamic properties without restricting the
Hilbert space in consideration. Thus, within the one-band
approximation, these expressions were valid for any temperature.
However, if the size of the system is finite, number fluctuations
$\Delta N$ must be included and to derive expressions valid for
arbitrary temperatures could be difficult. In this section we
calculate finite size corrections by restricting the temperature to
$kT <U/2$. At such temperatures
 Fig. \ref{fig1} shows that only states
with at most two atoms per site are relevant so one can restrict the
Hilbert space to include only states with at most two atoms per
site.

\begin{figure*}[htbh]
\begin{center}
\leavevmode {\includegraphics[width=7. in,height=3.2
in]{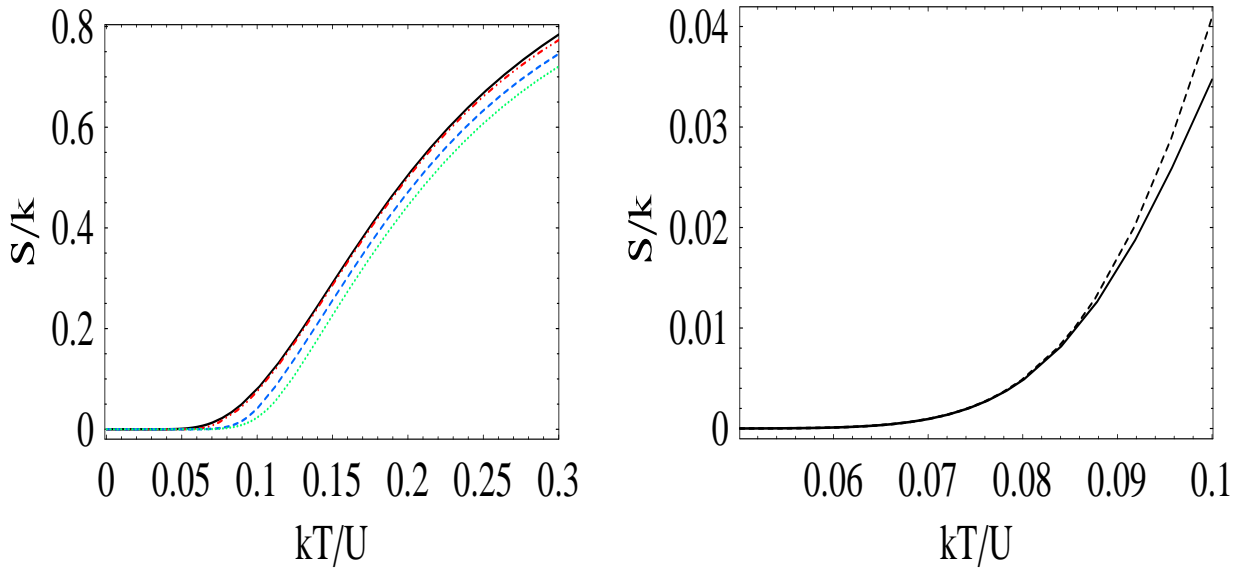}}
\end{center}
\caption{(color online) Left: Entropy per particle $S$ as a function
of the temperature $T$ (in units of U) for a unit filled lattice in
the $J=0$ limit and different number of atoms $N$.  The solid line
shows $S$ calculated in the thermodynamic limit using
Eq.(\ref{ana1}) and (\ref{ana2}). The dash-dotted(red),
dashed(blue), and dotted(green) lines correspond to $N=1000,100$ and
$50$, respectively. For these curves, $S$ is restricted to the ph
subspace (see Eq. (\ref{Zpatwo})). Right: $S$ vs $T$ (in units of U)
for $N=100$. The dashed and solid lines are the entropy calculated
in the ph and 1-ph(see Eq.\ref{S}) subspaces, respectively.
}\label{fig3}
\end{figure*}

Setting  $\bar{n}_{r>2}=0$ and $M=N$ in Eq.(\ref{Zpa}), the
partition function (at zero order in $J$ ) can be explicitly written
as:

\begin{eqnarray}
\mathcal{Z}(N,N)&=&\sum_{j=0}^{\lfloor
N/2\rfloor}\frac{N!}{(j!)^2(N-2j)!} e^{-\beta U j},\notag \\
&=&e^{-\beta U/2} \cos(\pi N) C_N^{(-N)}[\frac{1}{2} e^{\beta
U/2}],\label{Zpatwo}
\end{eqnarray}where $C_n^{(m)}[x]$ are Gegenbauer polynomials \cite{AS64}.

In Fig.\ref{fig3} (left panel) we study the effect of finite atom
number on the entropy. We show the entropy per particle as a
function of temperature for systems with $N=~50$ (green dotted
line), $N=~100$ (blue dashed line) and $N=~1000$ (red dash-dotted
line). For comparison purposes we also plot with a (black) solid
line  the entropy calculated using Eqs.(\ref{ana1}) and
(\ref{ana2}), which were derived in the thermodynamic limit. It can
be observed that for $N =1000$ the thermodynamic limit is almost
reached (nearly indistinguishable from the thermodynamic limit).
Finite size effects decrease the entropy per particle and thus tend
to increase the final temperature during the adiabatic loading.

 Furthermore, in the right panel  we also  compare  Eq. (\ref{Zpatwo})
with the entropy calculated
  by restricting the Hilbert space even more and including
only one-particle-hole
  (1-ph). 1-ph excitations are the lowest
lying excitations  which correspond to states that have one site
with two atoms, one with zero atoms and one atom in every other
site, i.e. $\{n_r\}_{U}=\{1,N-2,1,0,\dots,0\}$.  There are $ N(N-1)$
different particle hole excitations all with  energy $U$. If the
entropy is calculated taking into account only 1-ph excitations one
gets an expression to zeroth order in $J$ given by:

\begin{equation}
\frac{S}{k}\approx   \frac{\ln [1+N(N-1)e^{-\beta U}]}{N} +\frac{
\beta U (N-1)e^{-\beta U} }{1+N(N-1)e^{-\beta U}}. \label{S}
\end{equation}

 The right panel shows that as long as  the temperature is below $kT \ll 0.1 U$ and
the number of wells is of order $10^2$ or less, Eq.(\ref{S}) gives a
very good approximation for the entropy per particle.

\subsection{Finite J corrections}
 In the previous section for simplicity we
worked out the thermodynamic quantities assuming $J=0$. However, if
the final lattice is not deep enough, finite $J$ corrections should
be taken into account. In this section we study how these corrections can help to cool
the unit filled lattice during adiabatic loading.

In the $J=0$ limit  all thermodynamic quantities are independent of
the dimensionality of the system.  On the other hand, for finite $J$
the dimensionality becomes important. Including $J$ in the problem
largely  complicates the calculations as number Fock states are no
longer eigenstates of the many-body Hamiltonian and many
degeneracies are lifted. For simplicity, in our calculations we will
focus on the 1D case and  assume periodic boundary conditions. We
will also limit  our calculations to systems with less than $10^2$
atoms and temperatures low enough ($kT \ll 0.1 U$)   so it is
possible to restrict the Hilbert space to  include  only 1-ph
excitations.

To find first order corrections to the $N(N-1)$ low lying excited
states we must diagonalize the kinetic energy Hamiltonian within the
1-ph subspace. For 1-D systems this diagonalization yields   the
following approximated expression for the
eigenenergies\cite{AnaBragg}
\begin{equation}
E_{rR}^{(1)}=U-4J\cos \left( \frac{\pi r}{N}\right) \cos \left(
\frac{\pi R}{N}\right), \label{Spect}\end{equation} Where $r=1,\dots
N-1$ and $R=0,\dots N-1$. Using these  eigenenergies to evaluate the
entropy per particle  one obtains the following expression:
\begin{equation}
\frac{S}{k}\approx \frac{\rm{ln}Z}{N} + U\beta  (N-1)\frac{[
I_0^2(2J\beta )- \frac{4 J}{U} I_0(2J \beta)I_1(2J
\beta)]}{Z}e^{-\beta U},\notag \label{Sph}
\end{equation} with
\begin{equation}
Z=1+N(N-1)e^{-\beta U}I_0(2J \beta),
\end{equation} where $I_n(x)$ are modified Bessel functions of the
first kind \cite{AS64}.

To derive Eq.(\ref{Spect}), we assumed similar effective tunneling
energies for  the extra particle and the hole. This is not exact,
especially for a unit filled lattice, $g=1$, since the effective
hopping energy for the particles and holes  goes like $J(g+1)$ and
$J g$ respectively. However, we find by comparisons with the exact
diagonalization of the Hamiltonian that for observables such as the
partition function which involves summing over all the $1-$ph
excitations, this assumption compensates higher order corrections in
$J/U$ neglected to first order. It even gives a better expression
for  the  entropy of the many-body system than the one calculated by
using the spectrum obtained by exact diagonalization in the $1-$ph
subspace.

\begin{figure}[tbh]
\begin{center}
\leavevmode {\includegraphics[width=3.3 in,,height=2.7
in]{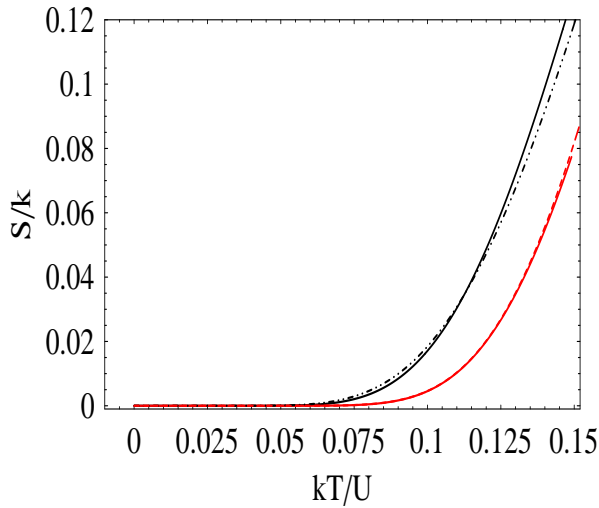}}
\end{center}
\caption{(color online) Finite $J$ corrections: The dash-dotted and
broken lines correspond to the entropy per particle vs T (in units
of U) calculated by numerical diagonalizations of the Hamiltonian
for systems with $N=10$ atoms and $J/U$=0.1 and 0.01 respectively.
The corresponding solid lines show the entropy per particle
calculated from Eq.(\ref{Sph}). }\label{fig4}
\end{figure}

\begin{figure*}[htbh]
\begin{center}
\leavevmode {\includegraphics[width=6 in]{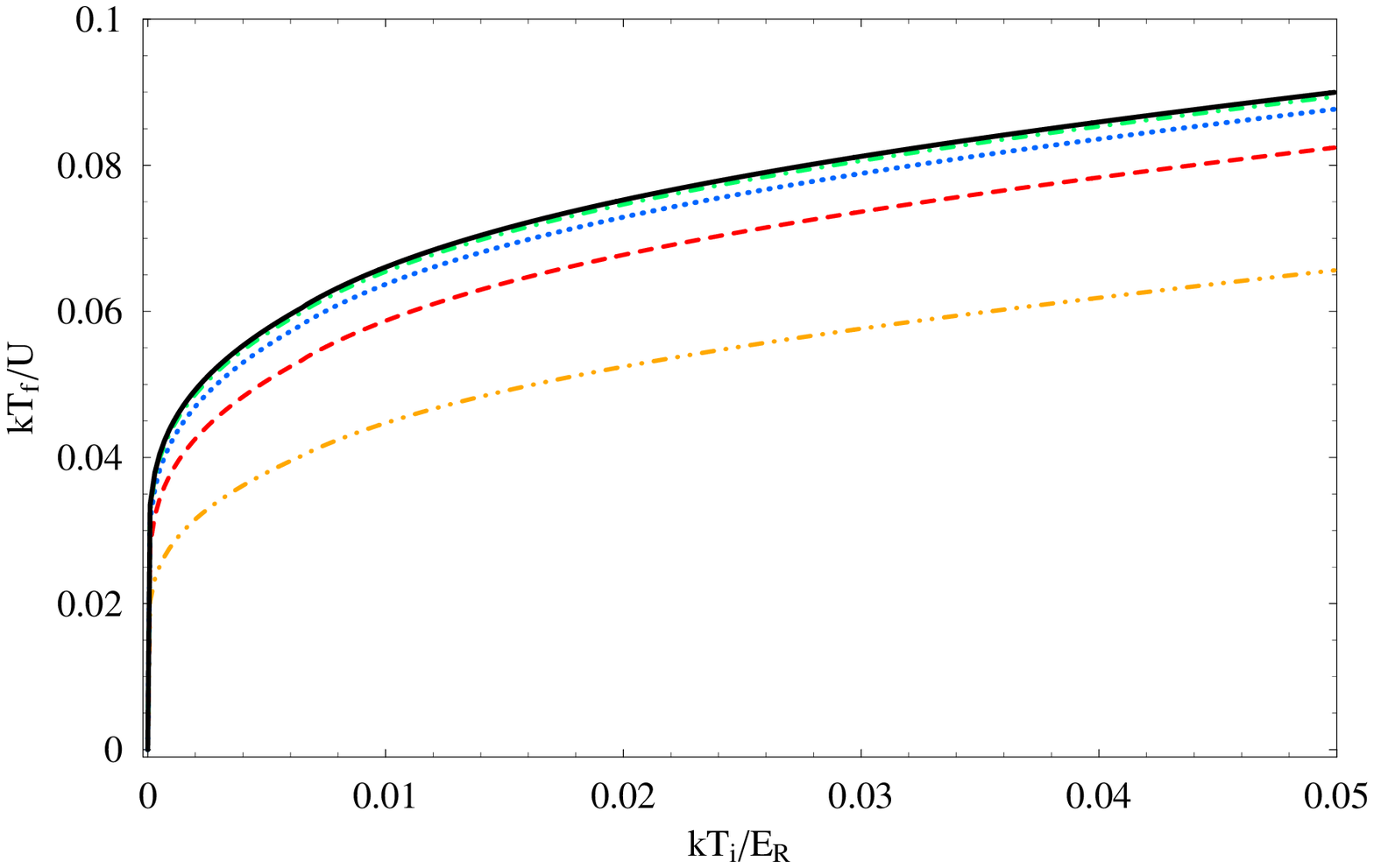}}
\end{center}
\caption{(color online) $T_f$ (in units of $U$) vs $T_i$ (in units
of $E_R$)  curves calculated using Eq.(\ref{Sph}) for a system with
$N=100$ atoms and different values values of $J/U$:
dash-dot-dotted(yellow) line $J/U=0.12(V_z= 5E_R)$, dashed(red) line
$J/U=0.07(V_z= 7E_R)$, dotted(blue) line $J/U=0.04(V_z= 9E_R)$ and
dash-dotted (green) line $J/U=0.02(V_z= 11E_R)$. The solid(black)
lines is shown for comparison purposes and corresponds to the
limiting case $J=0$.}\label{fig4a}
\end{figure*}

We show  the validity of Eq.(\ref{Sph}) in Fig.\ref{fig4} where we
compare its predictions ( plotted with solid lines)  with the
entropy calculated by exact diagonalization of the Bose-Hubbard
Hamiltonian for different values of $J/U$ assuming a system with
$N=10$ atoms. In the plot we use a dot-dashed line for $J/U=0.1$ and
a dashed line for $J/U=0.01$ . Even for the case $J/U=0.1$ we see
the analytic solution reproduces very well the exact solution,
especially at low temperatures. At $k T
>0.11U$ higher order corrections is $J/U$ become more important.

We now use Eq.(\ref{Sph}) to study larger systems where an exact
diagonalization is not possible. Even though we expect finite $J$
corrections to become important at lower temperatures   for larger
systems, we consider that for systems with less than $10^2$ atoms,
small values of $J/U$  and within the low temperature restriction,
Eq.(\ref{Sph}) can still give a fair description of the entropy. In
 Fig.\ref{fig4a} we show  the effect of finite $J$
corrections on the final temperature of a system of $100$
${^{87}}$Rb atoms  when adiabatically loaded. For the calculations,
we  fix the transverse lattice confinement to $V_x=V_y =30 E_R$,
assume  $d = 405$ nm and vary the axial lattice depth. We show the
cases  $V_z= 5 E_R$ with a yellow dash-dot-dotted line, $V_z= 7 E_R$
with a red dashed line, $V_z= 9 E_R$  with blue dotted line and
$V_z= 11 E_R $ with a green dash-dotted line. For these lattice
depths, the single-band approximation is always valid and the
energies $J$ and $U$ both vary so that their ratio decreases as
$J/U=\{0.12, 0.07, 0.04, 0.02\}$ respectively. We also plot for
comparisons purposes the $J/U=0$ case with a solid black line.

Fig.\ref{fig4a} shows that finite $J$ corrections decrease the final
available temperature of the sample. These  corrections are
important for shallower lattices, as they decrease the final
temperature with respect to the $J=0$ case by about $30 \%$. For
lattices deeper than  $V_z= 11 E_R $ the corrections are very small.

The  decrease in the final temperature induced by $J$  can be
qualitatively understood in terms of the modifications that hopping
makes to the eigenenergies of the system.  $J$ breaks the degeneracy
in the $1-$ph, leading to a quasi-band whose width is proportional
to $J$. As $J$ increases the energy of the lowest excited state
decreases accordingly,  while the ground state is only shifted  by
an amount proportional to $J^2/U$. The lowest energy excitations
then lie closer to the ground state and become accessible at lower
temperatures. As a consequence, the  the entropy increases (and thus
$T_f$ decreases) with respect to the $J=0$ case.

 Following the same lines of reasoning the entropy
 should exhibit a maximum at the critical point associated with the Mott insulator transition, since at
 this point an avoided crossing takes place. We confirmed this
 intuitive idea with exact numerical diagonalization of small
 systems. For the translationally invariant case, we expect the entropy to become sharply peaked at
the transition with increasing $N$ and  this could be an important
limitation for adiabatically loading atoms. However, as we will
discuss later, the external  harmonic confinement present in most
experiment prevents a sharp Mott insulator transition and can help
to decrease the adiabaticity loading time within accessible
experimental time scales.

In this section we focused on the effect of finite J corrections in
1D systems.  For higher dimensions, we expect that finite $J$
corrections help to cool the system  even more, since the effective
tunneling rate that enters in the entropy scales with the number of
nearest neighbors  and thus becomes larger for higher dimensions.

\section{Harmonic confinement: $V_i=\Omega i^2$ }
\label{para}

\begin{figure}[tbh]
\begin{center}
\leavevmode { \includegraphics[width=3.5 in,height=2.6
in]{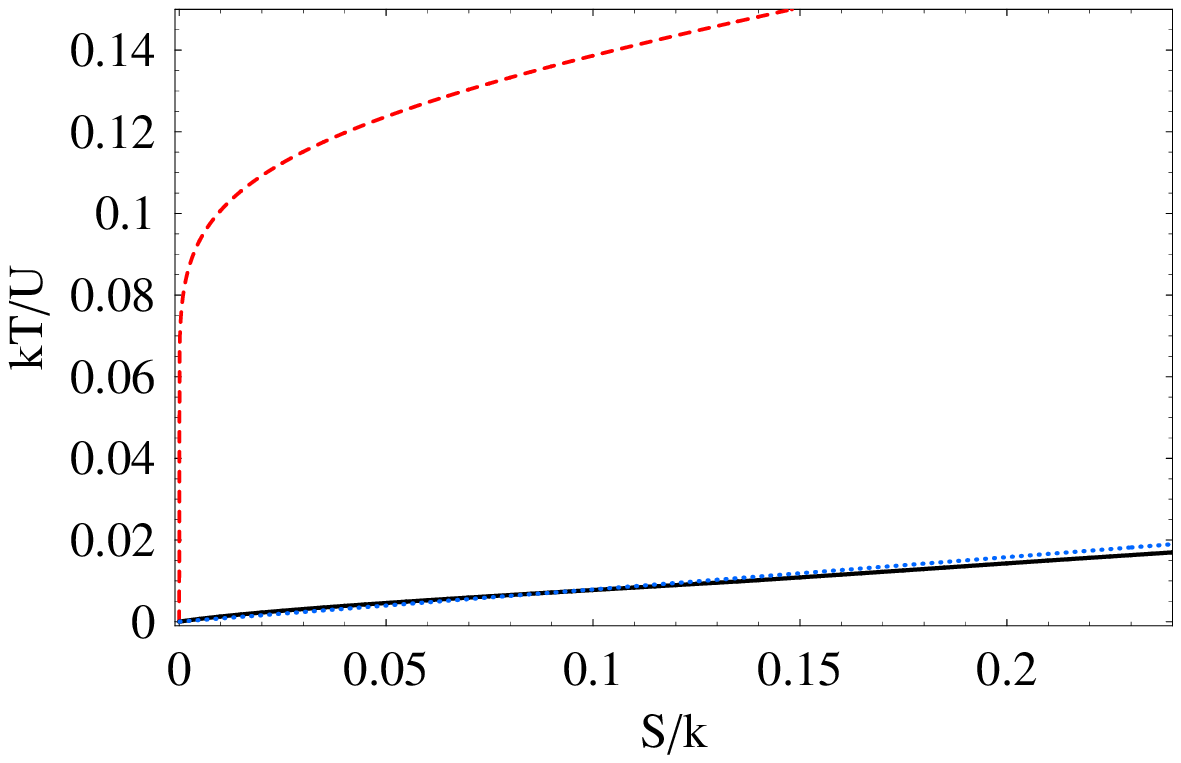}}
\end{center}
\caption{(color online) Final temperature (in units of U) vs initial
entropy per particle. The solid(black) line corresponds to the
trapped system with the parameters chosen to closely reproduce the
experimental set-up of Ref\cite{Paredes}. The dotted(blue) line is
the analytic solution for a system of fermions at low temperature
assuming a box-like spectrum and the dashed(red) line is the entropy
for the correspondent homogeneous system calculated from Eq.
(\ref{S})\label{fig5}}
\end{figure}

\begin{figure}[htbh]
\begin{center}
\leavevmode {\includegraphics[width=3.5 in,height=5.5in]{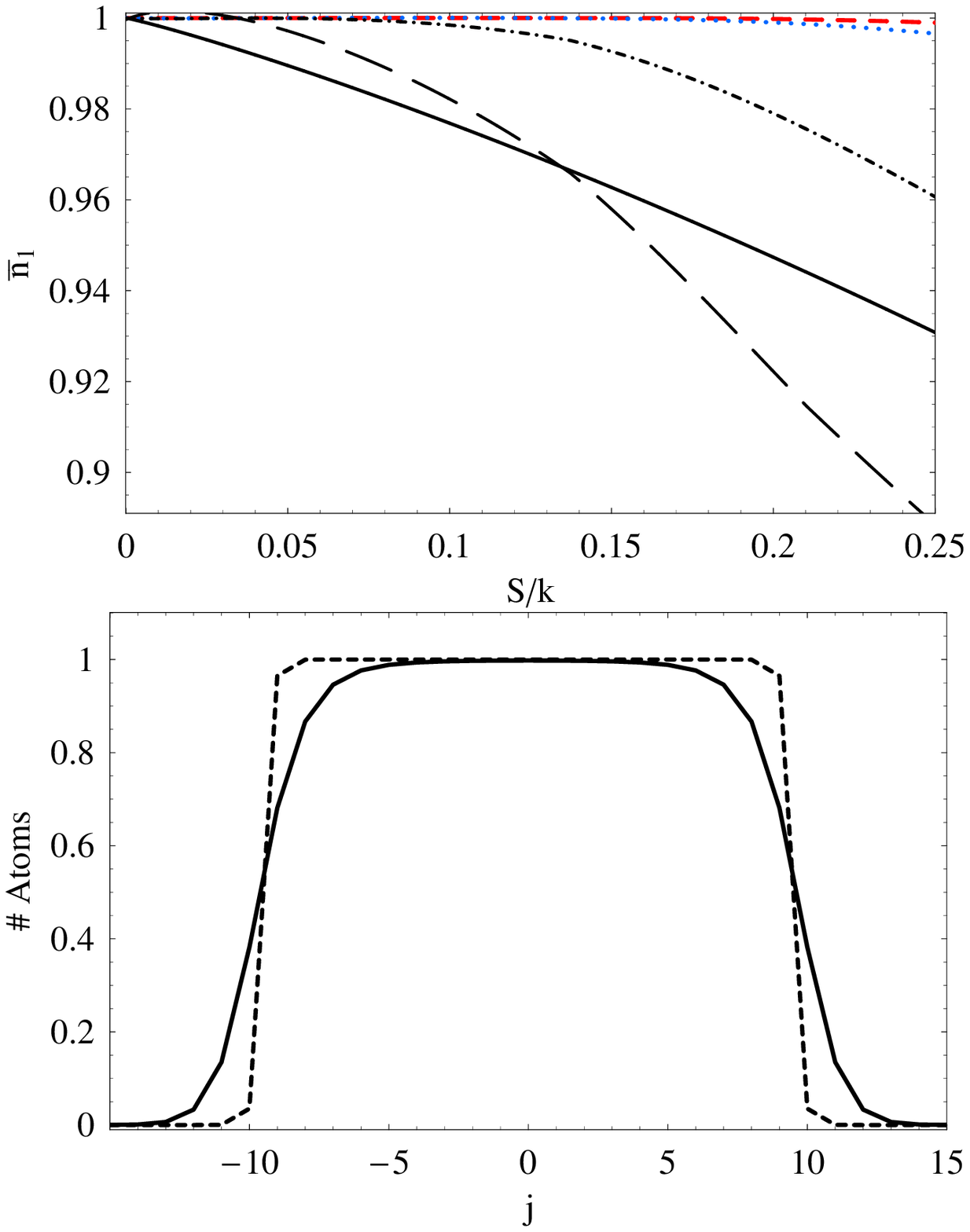}}
\end{center}
\caption{ (color online) Top: Local on-site probability of having
unit filling $\overline{n}_1$ as a function of the entropy per
particle $S$, and for a few values of the site-index $j$. The
dashed(red), dotted(blue), dot-dashed(black), and long-dashed(black)
lines correspond to the sites with $j=0,4,7$ and 8, respectively.
For comparison purposes, we also plot $\overline{n}_1$ calculated
for the correspondent spatially homogeneous system (solid line).
Bottom: Density profile for the trapped system in consideration at
$S/k=0 $ (dashed line) and $S/k=0.25$ (solid line). }\label{fig6}
\end{figure}

For simplicity in our analysis we consider a 1D system which can be
studied using  standard fermionization techniques \cite{GR}.  These
techniques allow us  to map the complex strongly correlated bosonic
gas into a non-interacting fermionic one. We choose our parameters
so that they closely resemble the experimental ones used in
Ref.\cite{Paredes}. Specifically we use  transverse lattice depths
of $V_x=V_y=27 E_R$ created by lasers with wavelength $\lambda_x=
823$ nm  and an  axial lattice depth of $V_z=18.5 E_R$ created by a
laser of wavelength $\lambda_z=854$ nm. We set the axial frequency
of the 1D gas   to $\omega_z=2 \pi \times 60$ Hz and the number of
atoms to $N=19$ (this was the mean number of atoms in the central
tube of the experiment). For these parameters the ratio $U/J\sim
205$ and $\Omega/J=0.28$ with $\Omega \equiv 1/8 m \omega _z
^2\lambda_z^2$. The ground state of the system corresponds to a MI
with $N$ unit filled sites  at the trap center  (see Fig.\ref{fig6}
bottom panel). We compare the thermodynamic properties of this
system with the ones of a translationally invariant system in the MI
state, with the same number of atoms ($N=M=19$) and same ratio
$U/J$.
As we described in the previous section, for homogeneous systems the
finite $J$ corrections for the  $J/U\sim 0.005 $ ratio in
consideration are very small, and for temperatures below
$kT/U\lesssim 0.1$ they can be neglected. On the contrary, when the
parabolic confinement is present, taking into account the kinetic
and trapping energy corrections is crucial for a proper description
of the low temperature properties of the system. Unlike the
spatially invariant case, where the lowest lying excitations in the
MI phase are 1-ph excitations which have an energy cost of order
$U$, in the trapped case, within the parameter regime in
consideration, there are always lower-lying excitations induced by
atoms tunneling out from the central core and leaving holes inside
it. We refer to these excitations as  {\it n-hole} (n-h)
excitations. These ``surface'' n-h  excitations must be included in
the trapped system because of the reservoir of empty sites
surrounding the central core, which introduces an extra source of
delocalization.
For example the lowest lying hole excitations correspond to
the 1-h excitations created when a hole tunnels into one of the most
externally occupied sites. They have energy cost $\Omega N $, which
for the parameters in Ref.~\cite{Paredes} is $40$ times smaller than
$U$.

For a system in arbitrary dimensions, it is complicated to properly
include n-h excitations in the calculations of thermodynamic
properties. For 1D systems, however, the Bose-Fermi mapping allows
us to include them in a very simple way.
 Nevertheless, because fermionization techniques neglect multi-occupied wells in the system
we have to restrict the analysis to temperatures at which no
multiple occupied states are populated ($kT\lesssim 0.1U$, see also
Ref.\cite{GCP}). The results are shown in Fig.\ref{fig5}, where we
plot the final temperature of the sample as a function of a given
initial entropy $S$. In the plot we also show the results for the
corresponding translationally invariant system.

The most important observation is that instead of the sudden
temperature increase at $S=0$ (or flat S vs T  plateau induced by
the gap), in the trapped case  the temperature increases slowly and
almost linearly with $S$:

\begin{eqnarray}
S &\approx& A \left(\frac{T}{T_F}\right),\\
A &=&5  k\left(\frac{\pi}{6}\right)^2, \label{A}
\end{eqnarray}
with $T_F$ the  Fermi temperature. The linear behavior is
characteristic of low temperature degenerate Fermi gases and the
proportionality constant $A$ depends on the density of states of the
system. For this particular case, $A$ can be estimated assuming a
box-like dispersion spectrum, $E_n=\Omega n^2$. For the parameter
regime in consideration this assumption is valid for the modes close
to the Fermi energy,  which are the relevant ones at low temperature
(Ref.\cite{Ana2005}). Using this box-like spectrum it is possible to
show that for $\Omega<  k T\ll k T_F$ (where the first assumption
allows a semiclassical approximation) $A$ is given by Eq.(\ref{A}).
In Fig.\ref{fig5} the blue-dotted line corresponds to this linear
solution and  it can be seen that it gives a fair description of the
entropy in the low temperature regime. It is interesting to point
out that  the slower increase in entropy as a function of
temperature in the homogeneous system compared to the trapped one is
a particular effect induced by interactions.  In the non-interacting
case the opposite behavior is observed:  for an homogeneous system
$S^{h}\propto (T/T_c)^{D/2}$ whereas for the trapped system
$S^{\omega}\propto (T/T_c)^{D}$ so if $T<T_c$ then
$S^{h}>S^{\omega}$. Here $D$ is the dimensionality of the system and
$T_c$ the critical condensation temperature.

In a typical experiment the sample is prepared by forming a BEC in a
magnetic trap. Therefore  a good estimate of the initial entropy is
given by \cite{Dalfovo}:

\begin{equation}
S=k \left(4 \zeta(4)/\zeta(3) t^3+\eta \frac{1}{3} t^2(1 -
t^3)^{2/5}\right),
\end{equation} where $\eta=\alpha(N^{1/6} a_s/\overline{a}_{ho})^{2/5}$, with $\alpha=15^{2/5}[\zeta(3)]^{1/3}/2\approx
1.57$ and $\overline{a}_{ho}=\sqrt{\hbar/(m\overline{\omega}) }$ the
mean harmonic oscillator length
($\overline{\omega}=\sqrt[3]{\omega_x \omega_y \omega_z}$). The
parameter $\eta$
 takes into account the main corrections to the entropy due to  interactions. In the above equation  $t=T/T_c$ with
 $T_c=T_c^0(1-0.43 \eta^{5/2})$ the critical temperature
for condensation and  $T_c^0$  the critical temperature for the
ideal trapped gas $k T^0_c=\hbar
\overline{\omega}(N/\zeta(3))^{1/3}$. The term proportional to $\eta
^{5/2}$ accounts for the small shift in the critical temperature
induced by interactions. For typical experimental parameters $\eta$
ranges from 0.35 to 0.4.

If one assumes a $N=10^{5-6}$ atoms,
$\overline{\omega}/(2\pi)=60-120$ Hz and a very cool initial sample,
$t\backsim 0.2$, one obtains that in typical experiments the initial
entropy per particle of the system is not smaller than $S/k\gtrsim
0.1$. Fig.\ref{fig5} shows then that the reduction of the final
temperature during the adiabatic loading induced by the trap can be
significant. In turn, this suggests that the presence of the
magnetic confinement is going to be crucial in the practical
realization of schemes for lattice-based quantum computation.

To emphasize this point, in Fig.~\ref{fig6} (top panel) we plot the
mean occupancy of some lattice sites  as a function of the initial
entropy per particle.
It should be noted that for the number of atoms in consideration the
edge of the cloud at $T=0$ is at $j=(N-1)/2=9$. For comparison
purposes we also plot $\overline{n}_1$ calculated for the
correspondent spatially homogeneous system. The plots shows that for
the central lattice sites there is  almost $100\%$ fidelity to have
one atom per site for the range of initial entropies in
consideration. Fluctuations are only important at the edge of the
cloud and if one excludes these extremal sites the fidelity in the
trap case remains always higher than the fidelity in absence of the
trap. In the bottom panel we also show the density profile for $S=0$
and compare it with the one at $S/k= 0.25$. It is clear in the plot
that the central lattice sites remain almost with one atom per site.

The considerations made here are for adiabatic changes to the
lattice, and therefore represent a lower bound on the final
temperature, assuming the entropy is fixed.  How quickly the lattice
can be changed and remain adiabatic is a separate issue, but we
point out that for systems with finite number of atoms confined by
an external trap there is not a sharp superfluid/insulator phase
transition, which should relax the adiabaticity requirements when
passing through the transition region. A proper adjustment of the
harmonic confinement during the loading process could reduce the
time scales required for adiabaticity to be experimentally
realizable.

\section{Conclusions}
\label{concl}

In this paper we calculated  entropy-temperature curves for
interacting bosons in unit filled optical lattices  and we used them
to understand how adiabatic changes in the lattice depth affect the
temperature of the system.

For the uniform system, we have derived analytic expressions for the
thermodynamic quantities in the $J/U=0$ case and we used them to
identify  the regimes  wherein adiabatically changing the lattice
depth will cause  heating or cooling of the atomic sample in the
case of a unit filled lattice. We have shown that the heating is
mainly induced by the gapped excitation spectrum characteristic of
the insulator phase. By considering finite  size effects and  finite
$J$ corrections we have shown that the former leads to increased the
heating of the atoms, the latter tend to reduce it.

Finally, we have discussed the spatially inhomogeneous system
confined in a
 parabolic potential and  we have shown that the presence of the trap
 reduces significantly the final available temperature of the atoms due to the
low-energy surface excitations always present in  trapped systems.

The fact that the harmonic confinement turns out to be clearly a
desirable experimental tool for reducing temperature in the lattice
is an important finding which should be taken into account in the
ongoing experimental and theoretical efforts aimed at using the Mott
Insulator transition as a means to initialize  a register for
neutral atom quantum computation.

\noindent\textbf{Acknowledgments}

This work is supported in part by the Advanced Research and
Development Activity (ARDA) contract and the U.S. National Science
Foundation through a grant PHY-0100767. A.M.R. acknowledges
additional support by a  grant from  the Institute of Theoretical,
Atomic, Molecular  and Optical Physics at Harvard University and
Smithsonian Astrophysical observatory. G.P. acknowledges additional
support from the European Commission through a Marie-Curie grant of
the Optical Lattices and Quantum Information (OLAQUI) Project.

\bigskip
{\it{Note:}} While preparing this manuscript  we have learned of a
recent report by K. P. Schmidt {\it et al.} \cite{ Schmidt} which
partially overlaps with our present work.

\end{document}